\documentclass[twoside]{article}
\usepackage{jfla99}
\usepackage{times}
\usepackage{epsfig}

\usepackage{alltt}

\title{Aspects de la programmation d'applications Win32 avec un langage
fonctionnel} 
\author{Riccardo Pucella$^1$\\Erik Meijer$^2$\\Dino Oliva$^1$}
\titlehead{Programmation d'applications Win32}
\authorhead{Pucella, Meijer \& Oliva}
\affiliation{
$^1$ Bell Laboratories, Lucent Technologies\\
Murray Hill, USA\\
\texttt{\{riccardo,oliva\}@research.bell-labs.com}
\\[.1in] 
$^2$ Department of Computer Science, University of Utrecht\\
Utrecht, The Netherlands\\
\texttt{erik@cs.ruu.nl}
}

\begin{document}
\maketitle

\begin{abstract}
Un langage de programmation qui se veut utile doit \^etre capable
d'exprimer des programmes qui profitent des services et des m\'ecanismes
de communication support\'es par le syst\`eme d'exploitation. Nous examinons
dans cet article le probl\`eme de la programmation d'applications
Win32 dites ``natives'' sous le syst\`eme d'exploitation Windows avec le
langage fonctionnel Standard ML. Nous introduisons une infrastructure bas\'ee
sur le langage d'interfaces IDL et sur une interface aux fonctions
\'etrang\`eres minimaliste pour explorer le support de l'API Win32 et de la
technologie COM dans le contexte de Standard ML.
\end{abstract}

\section{Introduction}

L'utilit\'e d'un langage de programmation moderne ne se juge pas uniquement
\`a la lumi\`ere de la qualit\'e intrins\`eque du langage. Comme l'indique
Wadler \cite{Wadler98}, un langage doit de plus poss\'eder un
environnement de programmation capable d'exprimer des programmes qui
coexistent avec un mod\`ele d'application support\'e par le syst\`eme
d'exploitation. Ce mod\`ele d'application peut inclure des services
fournis par le syst\`eme d'exploitation, comme l'interface-usager et
la gestion d'\'ev\'enements, et aussi inclure des m\'ecanismes de
communication et des ``contrats'' entre diff\'erentes applications
sous un m\^eme syst\`eme et par-del\`a un r\'eseau. 

Dans cet article, nous examinons un tel probl\`eme en
d\'etail. Sp\'ecifiquement, nous \'etudions la programmation d'applications
dites \emph{natives Win32} sous un syst\`eme Windows avec le langage
fonctionnel Standard ML (SML) \cite{Milner97}. Nous nous concentrons sur la 
communication avec le syst\`eme d'exploitation via l'API Win32, et la
communication avec d'autres applications via l'interface COM (\emph{Components
Object Model}).

Plusieurs buts guident notre entreprise. Tout d'abord, nous voulons unifier
les diff\'erentes visions du support du syst\`eme d'exploitation Windows, en
fournissant une infrastructure pour exprimer ces diff\'erentes
visions. Nous englobons le travail effectu\'e par l'\'equipe du
langage de programmation Haskell \cite{PeytonJones98, 
Finne98}, c'est-\`a-dire le \emph{scripting} de composants COM \`a
partir d'un langage fonctionnel, et le 
compl\'etons en fournissant acc\`es \`a l'API Win32, n\'ecessaire pour 
le d\'eveloppement de programmes Windows. L'utilisation du langage SML
apporte des avantages notables compar\'e \`a l'utilisation de
Haskell dans ce contexte. Principalement, le syst\`eme de modules de SML permet
de capturer de fa\c{c}on abstraite la notion d'interface avec le syst\`eme
Win32, tout en permettant des implantations diff\'erentes pour
diff\'erents compilateurs. Notre second but, plus p\'edagogique, est la
pr\'esentation informelle de la m\'ethodologie moderne d'utilisation de
composants telle qu'implant\'ee dans l'environnement Windows.  Cet
article donne un survol de notre project, en montrant qu'avec les
outils appropri\'es, il est simple de fournir une interface utilisable 
vers l'API Win32, et montre comment le reste de l'infrastructure
Windows peut suivre. Un prototype du syst\`eme existe, et est
implant\'e avec le compilateur Standard ML of New Jersey \cite{Appel87}.

Nous imposons la contrainte majeure de garder la g\'en\'eration
d'applications Win32 enti\`erement en SML. Nous voulons \'eviter de
devoir programmer des aspects de l'application en C\footnote{Il peut \^etre
d\'esirable de programmer des parties de l'application en C, mais ce ne devrait
jamais \^etre n\'ecessaire.}, et nous irons m\^eme jusqu'\`a d\'esirer
\'eviter de requ\'erir un compilateur C. Ceci nous permet d'obtenir un 
syst\`eme relativement ferm\'e, simplifiant l'utilisation et
permettant une interactivit\'e toujours possible avec les
environnements SML courants.

Il est utile de remarquer que la plupart des remarques que nous allons
effectuer sur la communication avec le syst\`eme d'exploitation et
l'utilisation des m\'ecanismes de composants peuvent \^etre g\'en\'eralis\'ees,
notablement aux composants CORBA disponibles sur plusieurs syst\`emes. Nous
n'allons pas nous attarder sur de telles g\'en\'eralisations ici. De
m\^eme, ce travail se g\'en\'eralise imm\'ediatement \`a d'autres
langages, notablement OCAML \cite{UNSTABLE:Leroy96}.

Le plan g\'en\'eral de l'article est le suivant. Nous allons tout d'abord
d\'ecrire la structure g\'en\'erale de l'API Win32 et de la technologie COM,
ainsi que les diverses d\'erivations telles que OLE et ActiveX. Nous
d\'ecrivons ensuite une interface aux fonctions \'etrang\`eres
minimaliste, qui nous permettra de communiquer avec le syst\`eme
d'exploitation. Ensuite, nous parlerons d'IDL, le langage utilis\'e pour
d\'ecrire les interfaces de l'API Win32 et d'objets COM, et de l'outil
qui permet de g\'en\'erer du code SML \`a partir d'une description
IDL. Nous montrons comment interfacer l'API Win32 dans ce contexte, et
comment utiliser et cr\'eer des objets COM. Finalement, nous allons
discuter certaines g\'en\'eralisations que nous allons explorer dans
le futur.

\section{La programmation Win32}
\label{sec:win32}

L'API Win32 est l'ensemble des fonctions syst\`emes fournies par les
diff\'erents syst\`emes d'exploitation 32 bits de Microsoft\footnote{En
fait, Windows NT et Windows 95 implantent des sous-ensembles un tant soit
peu diff\'erents de l'API Win32. Ces diff\'erences ne nous
concernerons pas.}. Ces fonctions sont distribu\'ees dans des librairies
dynamiques (\emph{dynamic link libraries} ou \emph{dll}) qui forment
le coeur des syst\`emes d'exploitation Windows.

La fonctionalit\'e de l'API peut se d\'ecomposer en sept grandes classes:
\begin{itemize}
\item gestion des fen\^etres;
\item interface graphique;
\item contr\^oles communs;
\item acc\`es au \emph{shell};
\item services du syst\`eme;
\item services de r\'eseau; 
\item internationalisation;
\end{itemize}

Il n'est pas utile de fournir l'acc\`es \`a toutes ces fonctions. Par exemple,
la gestion de la m\'emoire est automatique en SML; la gestion des
entr\'ees/sorties est fournie par la \emph{Basis Library}, ainsi que les
services de r\'eseaux et certains services du \emph{shell}. Nous nous
concentrons ici sur les fonctions de gestion des fen\^etres, de l'interface
graphique et certaines fonctions de service. Les librairies qui nous
int\'eressent sont principalement \texttt{user32.dll}, \texttt{gdi32.dll} et
\texttt{kernel32.dll}.

L'API Win32 est intrins\`equement li\'e \`a une structure
particuli\`ere de programmes, que
nous revisons ici. Un programme Win32 se compose d'un point d'entr\'ee
appel\'e \texttt{WinMain}. Le r\^ole de cette fonction est d'initialiser
l'application, typiquement en enregistrant diff\'erentes classes de fen\^etres
utilis\'ees. L'enregistrement d'une fen\^etre n\'ecessite
la d\'efinition d'une fonction \emph{callback} qui va traiter les messages
address\'es \`a la fen\^etre. Cette fonction de traitement de messages est la
fonction principale de toute application Win32. Il peut exister
plusieurs fonctions de traitement de messages si plusieurs fen\^etres
sont expos\'ees par l'application --- une fonction par fen\^etre est
la r\`egle.

Plut\^ot que d'accumuler ici des d\'etails inint\'eressants, nous r\'ef\'erons
le lecteur \`a un ouvrage tel que \cite{Petzold96} pour un vision en
profondeur de la programmation Win32.

\subsection{COM}

La technologie moderne des syst\`emes d'exploitation Windows pour la
r\'eutilisation de code et la distribution de librairies converge
vers la technologie COM \cite{Microsoft95,Rogerson97}. COM est une
interface binaire, et 
repose sur un minimum de principes simples, permettant une expressivit\'e de
concept impressionante. 

L'id\'ee de base sous-jacente \`a COM est la notion d'une \emph{interface}
vers un objet. Une interface est une vue d'un composant, et correspond
simplement \`a un tableau de pointeurs de fonctions
(m\'ethodes). \'Etant donn\'e une interface et un objet COM, il est
possible de questionner l'objet pour savoir s'il supporte 
l'interface consid\'er\'ee. Si l'interface est support\'ee, un pointeur vers
l'interface est retourn\'e, et \`a travers ce pointeur il est possible
d'invoquer les m\'ethodes implant\'ees par l'interface.

L'identification d'objets COM et d'interfaces s'effectue \`a l'aide
d'iden\-ti\-fi\-ca\-teurs uniques (GUID): un identificateur de classe (CLSID) est
utilis\'e pour parler d'un objet COM en particulier, et un identificateur
d'interface (IID) pour parler d'une interface particuli\`ere. Il est important
de noter que le CLSID d'un objet COM fait partie de sa description
formelle. Installer un objet COM sur un syst\`eme ajoute son CLSID \`a la
liste maintenue dans le \emph{Registry}, et de m\^eme pour les interfaces
implant\'ees par l'objet. Utiliser un objet COM requiert le CLSID de
l'objet en question, ainsi qu'un IID correspondant \`a l'interface initiale
que l'on voudrait obtenir. Par exemple, la fonction Win32
\texttt{CoCreateInstance} prend un CLSID et un IID en arguments,
cr\'ee une instance de l'objet en question et retourne un pointeur
vers l'interface correspondante, si elle est support\'ee par l'objet. 

Une interface sp\'eciale, l'interface \texttt{IUnknown}, est l'interface de
laquelle toutes les interfaces h\'eritent, et doit \^etre support\'ee par tout
objet COM. Cette interface est d\'efinie ainsi en IDL, un langage pour 
sp\'ecifier les interfaces (voir Section \ref{sec:idl}):
\begin{alltt}\scriptsize
  interface IUnknown \{
    HRESULT QueryInterface ([in] const IID& iid,
                            [out,iid_is (iid)] void **ppv);
    unsigned long AddRef ();
    unsigned long Release ();
  \}
\end{alltt}
La fonction la plus int\'eressante est la premi\`ere. Les deux autres
fonctions ont trait \`a la gestion de la m\'emoire de l'objet. La fonction
\texttt{QueryInterface} nous permet, donn\'e une interface,
d'acc\'eder \`a une autre interface via son IID. Puisque toutes les
interfaces h\'eritent de \texttt{IUnknown}, toutes les interfaces fournissent
la fonction \texttt{QueryInterface}. 

Nous avons pratiquement termin\'e la description de la technologie COM, tout
au moins du point de vue de l'utilisateur. Il y a quand m\^eme beaucoup de
d\'etails \`a consid\'erer, notablement pour ce qui a trait \`a la gestion de
la m\'emoire et de la dur\'ee de vie des objets COM, ainsi que des conventions
\`a respecter, tels que l'id\'ee qu'effectuer un \texttt{QueryInterface} sur
une interface quelconque d'un objet donn\'e pour l'interface
\texttt{IUnknown} doit retourner le \emph{m\^eme} pointeur, permettant
l'utilisation de l'interface \texttt{IUnknown} 
comme identificateur utile pour comparer l'identit\'e d'objets COM. Le lecteur
int\'eress\'e par de tels d\'etails se devra de lire un manuel dans la veine
de \cite{Rogerson97}.

Les technologies dites avanc\'ees de Microsoft sont toutes b\^aties sur les
fondations fournies par les objets COM et la notion d'interface. Ces
technologies sont simplement des sp\'ecifications d'interfaces que des
composants doivent implanter pour interagir avec le syst\`eme. Nous
revisons ici de mani\`ere informelle les plus importantes,
c'est-\`a-dire DCOM, Automation, OLE et ActiveX.

\subsection{DCOM}

DCOM est une extension de COM qui permet aux serveurs d'objets COM de vivre
sur une autre machine du r\'eseau. Une communication de type RPC (\emph{Remote
Procedure Call}) permet d'interagir avec les objets ainsi cr\'ees.

\subsection{Automation}

La technologie Automation permet de communiquer de mani\`ere
interpr\'et\'ee et dynamique avec un objet COM. Un objet COM supporte
Automation s'il implante l'interface \texttt{IDispatch}:
\begin{alltt}\scriptsize
  interface IDispatch : IUnknown \{
    HRESULT GetTypeInfoCount ([out] UINT* pctinfo);
    HRESULT GetTypeInfo ([in] UINT iTInfo,
                         [in] LCID lcid,
                         [out] ITypeInfo **ppTInfo);
    HRESULT GetIDsofNames ([in] const IID& riid,
                           [in,size_is (cNames)] LPOLESTR* rgszNames,
                           [in] UINT cNames,
                           [in] LCID lcid,
                           [out, size_is (cNames)] DISPID* rgDispId);
    HRESULT Invoke ([in] DISPID dispIdMember,
                    [in] const IID& riid,
                    [in] LCID lcid,
                    [in] WORD wFlags,
                    [in,out] DISPPARAMS* pDispParams,
                    [out] VARIANT* pVarResult,
                    [out] EXCEPINFO* pExcepInfo,
                    [out] UINT* puArgErr);
  \}
\end{alltt}

Ces m\'ethodes permettent d'appeler certaines fonctions en fournissant
leur nom sous forme de cha\^\i{}ne de caract\`eres. La m\'ethode
\texttt{GetIDsofNames} transforme cette cha\^\i{}ne de caract\`eres en
DISPID, un entier correspondant \`a la fonction, si elle existe. Ce
DISPID est pass\'e \`a la m\'ethode \texttt{Invoke} avec les param\`etres \`a
passer \`a la fonction.  

Les fonctions que \texttt{Invoke} peut appeler (la
\emph{dispinterface} en langue COM), sont arbitraires et d\'ependent
de 
l'implantation de \texttt{Invoke} pour une interface donn\'ee. Un des
aspects le plus int\'eressant d'Automation concerne l'implantation
d'interfaces dites \emph{dual}, c'est-\`a-dire d'interfaces COM dont
les m\'ethodes peuvent \'egalement
\^etre invoqu\'ees de mani\`ere dynamique. Une interface \emph{dual} d\'erive
non pas de \texttt{IUnknown}, mais de
\texttt{IDispatch}. L'implantation de \texttt{Invoke} pour cette
interface est telle que sa dispinterface correspond aux m\'ethodes de
l'interface. Ceci permet d'acc\'eder aux m\'ethodes d'une interface de 
mani\`ere directe via les m\'ecanismes de COM ou interpr\'et\'ee via les
m\'ecanismes d'Automation.

Une des restrictions impos\'ees par Automation concerne le type des
param\`etres pass\'es par l'interface de \texttt{Invoke}. Les param\`etres
doivent \^etre de type \texttt{VARIANT}, une union des types de bases connus
du syst\`eme. Ceci implique que les fonctions invoqu\'ees par \texttt{Invoke}
doivent v\'erifier dynamiquement que leurs arguments ont un 
type correct. Il n'existe pas de moyen automatique pour v\'erifier que
tout est correct lors de l'appel de la fonction. Les librairies de
types (\emph{type libraries}) all\`egent ce fardeau, mais nous ne nous 
y attarderons pas dans cet article.

Notons enfin que Automation est le m\'ecanisme principal pour la
distribution de composants vers des langages tels que Visual Basic.

\subsection{OLE}

OLE est un ensemble d'interfaces COM standardis\'ees, ayant principalement
rapport \`a des notions d'interfaces-usager. Celles-ci incluent les notions
suivantes:  documents compos\'es, activation en place, \emph{drag and
drop}. OLE d\'efinit une interface \texttt{IDataObject} qui permet de
transf\'erer des informations entre applications, par exemple par le biais du
\emph{Clipboard}, ou de \emph{drag and drop}. 

Une application qui veut implanter le \emph{drag and drop} doit fournir au
syst\`eme d'exploitation un \emph{Drop Target Object} avec lequel le syst\`eme
va communiquer lorsque l'usager tra\^\i{}ne un ic\^one sur l'application. Cet
objet est un objet COM qui implante l'interface OLE \texttt{IDropTarget}, et
qui s'enregistre aupr\`es du syst\`eme via la fonction
\texttt{RegisterDragDrop} de l'API Win32. Lorsque un objet est
tra\^\i{}n\'e sur l'application, une m\'ethode de \texttt{IDropTarget}
de l'application est appel\'ee, qui se voit passer un objet
\texttt{IDataObject} repr\'esentant l'information tra\^\i{}n\'ee.

\subsection{ActiveX}

Le terme ``ActiveX'' est le terme le plus sur-employ\'e de la
litt\'erature,  apr\`es le terme ``visuel''. \emph{ActiveX}   
est une extension de la technologie OLE, 
permettant \`a des applications OLE d'interagir avec et par-del\`a
l'Internet. D'autre part, \emph{ActiveX Controls} est le nouveau nom de ce qui
\'etait auparavant appel\'e \emph{OLE Controls} , deriv\'es des contr\^oles
\emph{VBX} de Visual Basic. Ils permettent l'implantation de
contr\^oles, qui sont une g\'en\'eralization des fen\^etres-enfants fournies
par Win32, comme par exemple le contr\^ole d'\'edition (\emph{Edit Control}). 

Finalement, \emph{ActiveX Scripting} d\'enote la technologie qui permet \`a
une  application d'ex\'ecuter du code interpr\'et\'e en appelant un
\'evaluateur externe, tel qu'un interpr\'eteur JavaScript ou Visual
Basic.

\section{Une interface aux fonctions \'etrang\`eres}
\label{sec:ffi}

Interagir avec des fonctions fournies par le syst\`eme d'exploitation
n\'ecessite une interface aux fonctions \'etrang\`eres (\emph{foreign
function interface} ou FFI) qui nous permet de communiquer avec le
syst\`eme. Pour avoir une utilit\'e et une aise d'utilisation
maximale, la FFI devrait pouvoir communiquer de mani\`ere
enti\`erement dynamique avec le sys\`eme, c'est-\`a-dire sans
n\'ecessiter la g\'en\'eration de code C ou de code compil\'e
s\'epar\'ement. De  plus, pour nos besoins particuliers, 
(\'ecrire des callbacks et g\'en\'erer des objets COM) nous avons aussi
besoin de pouvoir transformer des fonctions SML en des pointeurs de
fonctions qui peuvent \^etre appel\'es de C.    

Nous pr\'esentons ici l'interface minimaliste qui est utilis\'ee pour un
prototype de notre syst\`eme, avec l'id\'ee qu'elle est assez simple pour
\^etre port\'ee sans trop de probl\`emes, puisqu'elle ne d\'epend que tr\`es
peu sur la repr\'esentation internes de donn\'ees dans le compilateur SML. 
\begin{alltt}\scriptsize
  structure Wffi : sig
    type addr = Word32.word
    type word = Word32.word
    type library

    val alloc : int -> addr
    val free : addr -> unit
    val offset : addr * int -> addr
    val store : addr * word list -> unit
    val read : addr * int -> word list

    val addrToFun : addr -> (word list -> word)
    val funToAddr : (word list -> word) -> addr

    val library : string -> library
    val function : library -> string -> addr
  end
\end{alltt}

L'id\'ee de base de l'interface est celle de repr\'esenter
l'information traversant l'interface sous une forme primitive:
une repr\'esentation en mots. Nous aurions pu aussi tout ramener sous
forme d'octets, mais \'etant donn\'e que les arguments des fonctions
de l'API qui nous concernent ont tous des grandeurs multiples de mots,
notre choix nous simplifie la vie. Une fonction \'etrang\`ere est une
fonction de type \texttt{word list -> word}, prenant comme
argument une liste de mots repr\'esentant les arguments de la
fonction. Cette repr\'esentation accomode le passage d'arguments de
grandeur de plus d'un mot (des structs, par exemple).

Le coeur de l'interface est la paire de fonctions \texttt{funToAddr}
et \texttt{addrToFun}, qui respectivement transforment une fonction de 
type \texttt{word list -> word} (une \emph{fermeture}) en pointeur
de fonction (une addresse), et transforment un pointeur de fonction en 
fonction de type \texttt{word list -> word}. En g\'en\'eral, ce type
d'implantation n\'ecessite une m\'ethode de g\'en\'eration de code-objet
\`a l'ex\'ecution (\emph{runtime code generation}), que nous empruntons \`a
l'une des FFI existantes de Standard ML of New Jersey
\cite{Huelsbergen95}.

Les fonctions \texttt{library} et \texttt{function} permettent de
charger une librairie dynamique, et d'en extraire des pointeurs de
fonction. Par exemple, pour acc\'eder \`a la fonction \texttt{ShowWindow} de
la librairie \texttt{user32.dll}:
\begin{alltt}\scriptsize
  val showWindow = let 
    val l = Wffi.library ("user32.dll")
    val f = Wffi.function (l,"ShowWindow")
  in
    Wffi.addrToFun (f)
  end
\end{alltt}

Il y a plusieurs points importants que nous n'abordons pas ici. Tout
d'abord, le fait qu'il existe plusieurs conventions d'appel (\emph{calling
conventions}) possibles, qui d\'eterminent si l'appelant ou l'appel\'e
est en charge de d\'epiler les arguments lorsqu'un appel de fonction
se termine. Nous ne ferons que noter que (presque) toutes les
fonctions de l'API Win32 et COM sont implant\'es avec la convention
dites \emph{Pascal}, qui dit que l'appel\'e d\'epile les arguments. Un
autre point que nous n'allons pas discuter concerne l'ordre dans
lequel les \'el\'ements de structures doivent \^etre empil\'ees. Ces
d\'etails, aussi importants soient-ils, ne feraient que compliquer
inutilement la pr\'esentation.

\section{Un langage de d\'efinition d'interfaces}
\label{sec:idl}

Notre but premier \'etant de fournir \`a l'utilisateur du langage SML une
fa\c{c}on d'acc\'eder aux fonctions fournies par l'API Win32 et par divers
objets COM, il est utile d'identifier une notation qui nous permette
de sp\'ecifier l'interface d'un syst\`eme d'une mani\`ere
ind\'ependante de la FFI du compilateur SML que nous
avons sous la main, et de r\'ealiser un outil qui nous permettrait de
convertir une description dans cette notation en une implantation
correspondant \`a la FFI qui nous int\'eresse.  L'utilisation d'une
FFI n\'ecessitant g\'en\'eralement du verbiage important et \'etant
susceptible \`a des erreurs difficiles \`a retracer, cette m\'ethode a
l'avantage de centraliser dans l'outil la g\'en\'eration de code de
support pour l'appel de fonctions \'etrang\`eres, ainsi que toute
d\'ecision portant par exemple sur la repr\'esentation des donn\'ees
(types de base, types construits tels que structures and unions).

La notation que nous adoptons est naturelle, et Microsoft
l'utilise d\'ej\`a pour d\'ecrire l'interface aux objets COM. Il s'agit
d'IDL (\emph{Interface Definition Language}), un langage d\'eriv\'e du mod\`ele
IDL de \cite{Xopen93} utilis\'e par l'OSF pour sp\'ecifier des services
RPC. IDL permet de d\'efinir des structures de donn\'ees semblables \`a celles
de C, et permet de d\'efinir des op\'erations d'une mani\`ere similaire \`a
C. Une diff\'erence importante concerne le fait que chaque param\`etre d'une
op\'eration poss\`ede un attribut qui d\'ecrit si le param\`etre est \emph{in}
(le param\`etre est utilis\'e pour passer une valeur \`a l'op\'eration), ou
\emph{out} (le param\`etre est utilis\'e pour retourner une valeur de
l'op\'eration). L'IDL que l'outil accepte est en fait une version
r\'eduite de l'IDL pour COM: la base de DCE, sans les attributs pour
la distribution par RPC, et la base de COM et Automation.

Voici un example de description IDL pour une librairie hypoth\'etique:

\begin{alltt}\scriptsize
  interface Time \{

    typedef struct \{
       long	sec;
       long	usec;
    \} timeval_t;

    void gettime ([out] timeval_t *cpu, 
                  [out] timeval_t *gc, 
                  [out] timeval_t *sys);

    void timeofday ([out] timeval_t *tod);
  \}
\end{alltt}

\begin{figure}[t]
\begin{center}
\mbox{\epsfxsize=200pt\epsfysize=217.5pt\epsffile{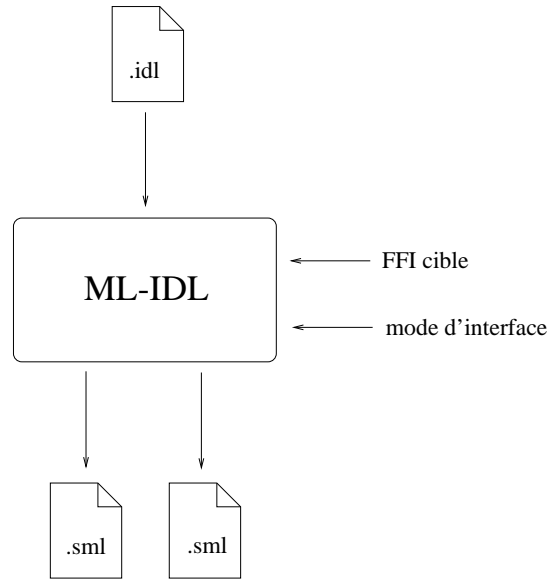}}
\end{center}
\caption{L'outil ML-IDL}
\label{fig:mlidl}
\end{figure}

Le mode de fonctionnement de notre outil, ML-IDL \cite{UNSTABLE:Pucella98},
est repr\'esent\'e \`a la Fig. \ref{fig:mlidl}. Comme nous l'avons dit, c'est
un traducteur de description IDL en code SML. L'outil est param\'etr\'e par
le type de FFI pour laquelle le code doit \^etre g\'en\'er\'e, et par le mode
d'interface d\'esir\'e --- un fichier IDL peut d\'ecrire une librairie statique,
une librairie dynamique, une interface COM. Le code g\'en\'er\'e repose sur la
capacit\'e d'abstraction du syst\`eme de modules de SML. Une signature et un
module sont g\'en\'er\'es pour chaque interface. La signature est
ind\'ependante de la FFI vis\'ee, et repr\'esente l'interface
abstraite. Le module g\'en\'er\'e varie selon la FFI vis\'ee.

L'outil ML-IDL est de plus param\'etr\'e par ce que l'on pourrait
appeler un ``niveau'' d'interface. L'id\'ee de base est de
repr\'esenter l'interface vers la librairie d\'ecrite en IDL de
mani\`ere abstraite, en utilisant des types abstraits correspondant
aux types de base et aux types d\'efinis, avec des fonctions cr\'eant
et lisant ces valeurs abstraites, et des fonctions agissant sur ces
valeurs abstraites correspondant aux fonctions de la librairie. Il est 
possible d'interagir avec la librairie de cette mani\`ere, en
convertissant ``\`a la main'' les valeurs SML \`a passer \`a la
librairie. Il est aussi possible de g\'en\'erer automatiquement ces
conversions, \`a la mani\`ere de H/Direct \cite{Finne98}. Mais
l'utilisation d'une interface abstraite permet d'implanter
diff\'erents niveaux de g\'en\'eration automatique, par exemple en
convertissant des valeurs abstraites retourn\'ees par la librairie de
mani\`ere paresseuse. L'outil ML-IDL peut donc \^etre vu comme un
g\'en\'eralisation de H/Direct dans ce contexte, en ce qu'il permet de 
multiples modes de traduction. Une description compl\`ete de l'outil
\'etant impossible, nous r\'ef\'erons le lecteur vers
\cite{UNSTABLE:Pucella98}. 

Pour les besoins de cet article, nous consid\'erons une traduction
d'IDL dans le style de H/Direct, qui nous donne pour la description
IDL donn\'ee plus haut la signature suivante. Noter que les
param\`etres \emph{out} sont retourn\'es comme r\'esultat
d'op\'erations. 

\begin{alltt}\scriptsize
  structure Time : sig
    datatype timeval_t = timeval_t of \{sec : Int32.int, 
                                       usec : Int32.int\}
    val gettime : unit -> (timeval_t * timeval_t * timeval_t)
    val timeofdat : unit -> timeval_t
  end
\end{alltt}

\section{Support pour l'API Win32}

L'API Win32 \'etant fixe, cr\'eer une implantation de l'API devient 
simplement une question de produire sa description IDL. Avec cette
description, nous pouvons utiliser l'outil ML-IDL pour g\'en\'erer une
implantation de l'API compatible avec la FFI du syst\`eme utilis\'e. 

L'appendice \ref{app:idl} pr\'esente un extrait de la description IDL
de l'API Win32 que nous utilisons. L'appendice \ref{app:sig}
pr\'esente la signature g\'en\'er\'ee par ML-IDL correspondant \`a
l'extrait. 

Nous ne montrons pas le module g\'en\'er\'e correspondant
\`a l'implantation de la signature par la FFI de la section
\ref{sec:ffi} pour des raisons d'espace. Plut\^ot qu'aller dans les
d\'etails, notons quelques aspects de la programmation Win32 en
SML par le biais d'un exemple. L'appendice \ref{app:exemple} pr\'esente un
programme Win32 \'ecrit en SML via l'interface de l'API Win32. Il
s'agit du logo de SML/NJ rebondissant \`a l'int\'erieur d'une
fen\^etre. Le code, une simple traduction du programme C
correspondant, montrent certaines caract\'eristiques de la
programmation Win32 en SML. 

Comme nous l'avons indiqu\'e \`a la Section \ref{sec:win32}, la
structure d'un programme inclut la d\'efinition de fonctions en charge
de traiter les messages address\'es aux diff\'erentes fen\^etres de
l'application. Nous appelons ces fonctions des wndprocs. De telles
wndprocs sont utilis\'ees lors de l'enregistrement de la classe d'une
fen\^etre. Une wndproc est un des \'el\'ements de la struct pass\'ee
comme argument \`a la fonction
\texttt{W32.User.RegisterClassExA}. L'outil ML-IDL g\'en\`ere les
abstractions n\'ecessaires pour nous permettre de passer une fonction
de type  
\begin{verbatim}
W32.hwnd * int * Word32.word * Word32.word -> int
\end{verbatim}
comme wndproc \`a la fonction \texttt{RegisterClassExA}. \'Etant
donn\'e que la struct  pass\'ee comme argument \`a la  
fonction \texttt{RegisterClassExA} de l'API s'attend \`a recevoir un pointeur vers
une fonction comme \'el\'ement correspondant \`a la wndproc, notre code SML
g\'en\'er\'e se doit de convertir la fonction SML en pointeur de
fonction. Ceci est achev\'e dans notre FFI en utilisant la fonction
\texttt{Wffi.funToAddr}.

Une cons\'equence de l'utilisation d'une wndproc comme structure centrale
d'un programme Win32 est particuli\`erement ennuyante. La wndproc est
appel\'ee une fois pour chaque message re\c{c}u de la fen\^etre
correspondante. Ce qui veut dire que si des donn\'ees doivent \^etre
pr\'eserv\'ees d'un message \`a l'autre, par exemple des donn\'ees sur les
dimensions de la fen\^etre, une cellule \texttt{ref} doit \^etre allou\'ee
pour l'information. Ce qui conduit \`a du code extr\^emement imp\'eratif, avec
beaucoup de variables globales. L'argument n'est pas si grave pour Standard ML
qui poss\`ede un support respectable pour la programmation imp\'erative, mais
rend l'expression de programmes Win32 natifs en Haskell (ou tout langage
fonctionnel pur) probl\'ematique. 

Il est possible de contourner ce probl\`eme en utilisant Concurrent ML
\cite{Reppy91}. Nous pouvons exprimer notre wndproc de la fa\c{c}on
suivante:
\begin{alltt}\scriptsize
  local
    val ch = channel (W32.nullHwnd, 
                      0,
                      0w0 : Word32.word, 
                      0w0 : Word32.word)
    val retCh = channel (0)
  in
    fun WndProc (hwnd, msg, wparam, lparam) = 
           (send (ch,(hwnd,msg,wparam,lparam));
            recv (retCh))

    fun WndProc' () = let
      val loop (state) = let
        val (hwnd,msg,wparam,lparam) = recv (ch)
      in
        (* body *)
      end
    in
      loop (initial_state)
    end
  end
\end{alltt}
Le corps de la boucle interne de \texttt{WndProc'} doit communiquer la
valeur de retour par le canal \texttt{retCh}, et s'appeler r\'ecursivement
avec un nouvel \'etat (contenant, par exemple, les dimensions de la fen\^etre
courante). Le code d'initialisation de l'application doit appeler
\texttt{spawn (WndProc')} pour cr\'eer la \emph{thread} qui va \'ecouter sur
le canal \texttt{ch}. 

Il ne reste qu'une question: pourquoi nous attardons-nous \`a vouloir
programmer \`a ce niveau si bas? Apr\`es tout, la plupart des programmeurs
d'applications Win32 ne programment pas directement avec l'API Win32, mais
plut\^ot avec une librairie de classes C++ telle que MFC ou OWL. En
fait, la question se pose de m\^eme en C++: pourquoi devrait-on
conna\^\i{}tre l'API Win32 s'il existe des librairies  comme MFC et
OWL? La r\'eponse dans les deux cas est similaire: parce ce qu'aucun
vernis de haut-niveau ne peut traiter tous les cas possibles. Si
quelque chose de sp\'ecial doit \^etre effectu\'e, il 
est souvent n\'ecessaire de revenir au niveau le plus bas, c'est-\`a-dire
l'API Win32. Il est donc imp\'eratif de pouvoir fonctionner compl\`etement \`a
ce niveau. Il est aussi int\'eressant de fournir une interface \`a un
niveau tel qu'il est possible d'utiliser directement des textes
p\'edagogiques existant. Il est en effet facile de traduire des
programmes tir\'es de \cite{Petzold96} et de les implanter en SML avec 
l'interface d\'eriv\'ee plus haut. Par exemple, l'exemple de
l'appendice \ref{app:exemple} est tir\'e directement du chapitre 7 de
\cite{Petzold96}.

\section{Support pour COM}

Il  a deux probl\`emes relativement distincts reli\'es \`a la programmation 
COM: utiliser des objets COM lors de la programmation d'applications Win32, et
l'implantation d'objets COM utilisable par d'autres applications. Nous
consid\'erons ces probl\`emes s\'epar\'ement. 

\subsection{Utiliser des objets COM en SML}

Le probl\`eme le plus simple \`a r\'esoudre est celui d'utiliser des objets
COM dans un programme SML. \'Etant donn\'e un objet COM que nous voulons
utiliser avec sa description IDL, nous pouvons utiliser l'outil ML-IDL pour
g\'en\'erer du code qui va pouvoir cr\'eer des instances de cet objet. 

Nous avons tout d'abord besoin d'une petite librairie COM fournissant
l'infrastructure n\'ecessaire. Les fonctions de cette librairie se retrouvent
principalement dans la librairie dynamique \texttt{ole32.dll}, mais nous ne
voulons pas utiliser la m\'ethode de la section pr\'ec\'edente pour
implanter cette interface, puisque nous voudrions un syst\`eme de type
plus sophistiqu\'e pour traiter les interfaces COM. Voici la signature d'une
partie de la librairie COM dont nous avons besoin. 

\begin{alltt}\scriptsize
  structure Com : sig

    type CLSID
    type 'a IID
    type 'a interface

    type IUnknown
    val IUnknown : IUnknown IID

    val CoCreateInstance : CLSID -> 'a IID -> 'a interface

  end
\end{alltt}

La fonction \texttt{CoCreateInstance} prend comme argument le CLSID de l'objet
que nous voulons cr\'eer, et l'interface que nous d\'esirons obtenir. 
Nous utilisons des types-t\'emoins (comme dans \cite{PeytonJones98}) pour
guarantir le fait que \texttt{CoCreateInstance} cr\'ee une interface (de type
\texttt{interface}) correspondant \`a l'\texttt{IID} demand\'e. Chaque interface
g\'en\'er\'ee par le compilateur IDL se doit d'inclure un type-t\'emoin pour
l'interface.   

Par exemple, consid\'erons un objet COM \texttt{Bar} qui poss\`ede une
interface \texttt{IX} et \texttt{IY} (toutes deux d\'eriv\'ees de
\texttt{IUnknown}) avec une fonction \texttt{FooX} dans \texttt{IX} et une
fonction \texttt{FooY} dans \texttt{IY}. La signature g\'en\'er\'ee par
l'outil IDL a la forme suivante. 

\begin{alltt}\scriptsize
  structure Bar : sig

    val BarCLSID : Com.CLSID

    structure IX : sig
      type IX
      val IX : IX Com.iid

      val QueryInterface : IX Com.interface -> 'a Com.IID -> 'a Com.interface
      val FooX : IX Com.interface -> unit -> unit
    end

    structure IY : sig
      type IY
      val IY : IY Com.iid

      val QueryInterface : IY Com.interface -> 'a Com.IID -> 'a Com.interface
      val FooY : IY Com.interface -> unit -> unit
    end

  end
\end{alltt}

Notons que les fonctions \texttt{AddRef} et \texttt{Release} de l'interface
\texttt{IUnknown} ne sont pas incluses dans le code g\'en\'er\'e, puisqu'elles
ne sont jamais utilis\'ees par l'utilisateur dans l'implantation SML. 

Dans notre prototype utilisant la FFI d\'ecrite \`a la Section \ref{sec:ffi},
une interface est simplement un pointeur vers un bloc de m\'emoire contenant
les pointeurs de fonctions. Extraire une m\'ethode d'une interface correspond
donc simplement \`a extraire le pointeur correspondant et de convertir le
pointeur de fonction en une fermeture via \texttt{Wffi.addrToFun}.

\begin{alltt}\scriptsize
  type 'a interface = Wffi.addr

  fun getMethod (interface,index) = let
    val i' = Wffi.offset (interface, index)
    val p = hd (Wffi.read (i',1))
  in
    addrToFun (p)
  end
\end{alltt}

\subsection{Cr\'eation d'objets COM en SML}

\`A priori, c'est le probl\`eme le plus ardu. Mais la difficult\'e est
principalement due \`a l'architecture du compilateur que nous utilisons. Le
probl\`eme peut se subdiviser encore une fois en deux probl\`emes
distincts. Tout d'abord, nous devons pouvoir cr\'eer des objets COM en
SML. Ensuite, nous devons pouvoir distribuer ces objets COM pour que d'autres
applications puissent les utiliser (programmation d'un serveur). 

Cr\'eer des objets COM est fondamentalement simple, mais demande de la
discipline. Comme on se le rappelle, un objet COM est simplement un
ensemble d'interfaces, toutes deriv\'ees de l'interface
\texttt{IUnknown}. Cr\'eer un objet COM correspond \`a determiner un ensemble
d'interfaces \`a supporter, implanter les m\'ethodes, et cr\'eer les
interfaces elles-m\^emes, en cr\'eant un tableau en m\'emoire et en y
pla\c{c}ant les pointeurs de fonctions correspondants. Comment construire une interface?
Supposons que nous voulons implanter l'objet COM \texttt{Bar} d\'ecrit \`a
la sous-section pr\'ec\'edente. Voici un extrait de code utilisant notre
FFI de la section \ref{sec:ffi}.

\begin{alltt}\scriptsize
  fun fooX [] = print "executing FooX"

  fun fooY [] = print "executing FooY"

  fun makeInterface (l) = let
    val l' = [queryInterface,addRef,release]@l
    val i = Wffi.alloc (length (l'))
    val _ = Wffi.store (i, map Wffi.funToAddr l')
    val p = Wffi.alloc (1)
    val _ = Wffi.store (p,[i])
  in
    p
  end

  val IUnknown = makeInterface []

  val IX = makeInterface [fooX]

  val IY = makeInterface [fooY]
\end{alltt}
La fonction \texttt{makeInterface} prend une liste de fonction SML \`a
ins\'erer dans l'interface, et construit le bloc m\'emoire, incluant les
fonctions faisant partie de l'interface \texttt{IUnknown}. La fonction
\texttt{queryInterface},  non montr\'ee, est en charge de d\'eterminer quelle
interface retourner (\texttt{IUnknown}, \texttt{IX} ou \texttt{IY}),
selon l'IID pass\'e.

La cr\'eation d'un serveur, c'est-a-dire d'un distributeur d'objets COM,
est plus compliqu\'ee. Tout d'abord, une usine de classes (\emph{class factory})
doit \^etre implant\'ee. Il s'agit l\`a d'un objet COM ayant pour fonction de
cr\'eer les objets COM implant\'es par le composant. Il n'y a pas
d'interface exig\'ee pour une usine de classes, mais l'interface
typique est l'interface \texttt{IClassFactory}. Il existe aussi une interface
\texttt{IClassFactory2} qui supporte des licenses et des
authorisations. Un client qui utilise des objets COM fournis par un serveur 
implantant une usine de classes d\'erivant de \texttt{IClassFactory} peut
utiliser la fonction \texttt{CoCreateInstance} de l'API Win32 pour cr\'eer les
objets directement. Si une autre interface est utilis\'ee, le client doit
utiliser la fonction \texttt{CoGetClassObject} de l'API Win32, qui extrait
l'usine de classes du composant et laisse au client le soin d'appeler la
fonction de cr\'eation requise. Il faut noter que \texttt{CoCreateInstance}
est implant\'e en fonction de \texttt{CoGetClassObject}.  

\begin{figure}[t]
\begin{center}
\mbox{\epsfxsize=145pt\epsfysize=154pt\epsffile{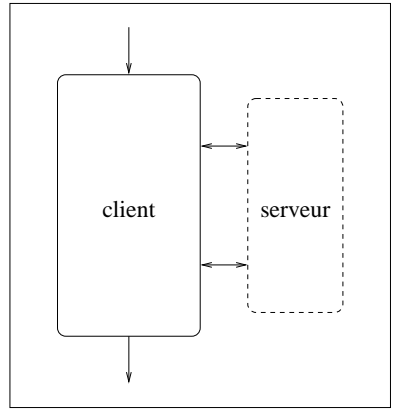}}
\end{center}
\caption{Serveur \emph{in-proc}}
\label{fig:inproc}
\end{figure}

Deux types de serveurs sont possible: \emph{in-proc} et
\emph{out-proc}. Un serveur \emph{in-proc} (voir Fig. \ref{fig:inproc})
est une librairie dynamique qui implante le code des objets COM
fournis par le composant. Une fonction \texttt{DllGetClassObject} doit \^etre
export\'ee par la librairie, et cr\'ee l'usine de classes pour le composant
(c'est la fonction qui est appel\'ee par \texttt{CoGetClassObject}). D'autres
fonctions, telles que \texttt{DllRegisterServer}, \texttt{DllUnregisterServer}
et \texttt{DllCanUnloadNow} doivent \^etre implant\'ees pour
l'enregistrement et la lib\'eration de la librairie dynamique. Le probl\`eme
de cette approche dans notre cas est que cr\'eer des librairies dynamiques
avec Standard ML of New Jersey n'est pas r\'eellement support\'e. Un
nouveau \emph{runtime system} est en d\'eveloppement qui permettra la
cr\'eation de librairies dynamiques et donc de serveurs in-proc.

\begin{figure}[t]
\begin{center}
\mbox{\epsfxsize=307pt\epsfysize=154pt\epsffile{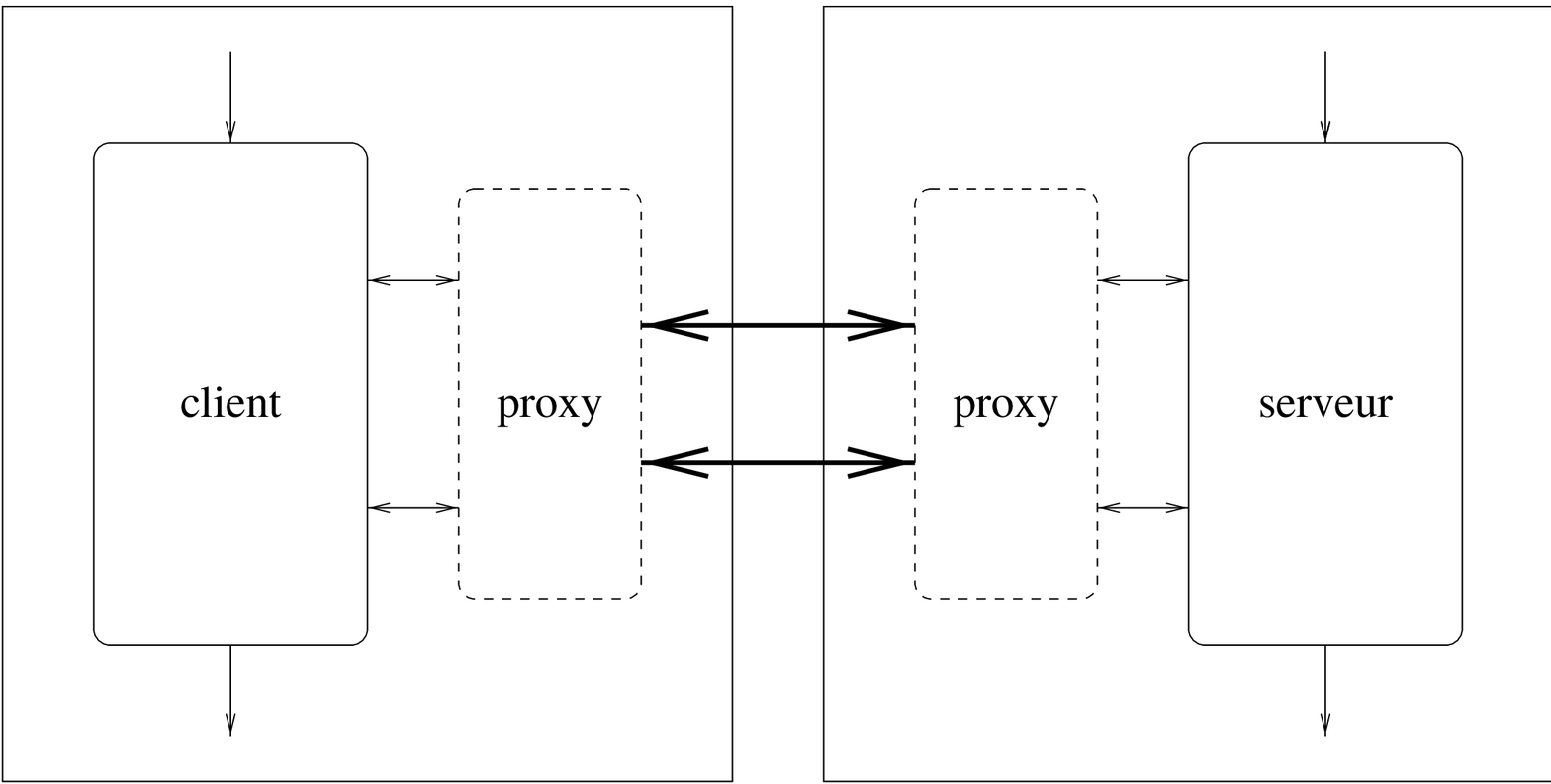}}
\end{center}
\caption{Serveur \emph{out-proc}}
\label{fig:outproc}
\end{figure}

L'autre type de serveur, un serveur \emph{out-proc}, est un serveur
implant\'e par un ex\'ecutable (un \emph{exe}). Si une librairie
dynamique fournissant un composant se doit d'exporter une fonction
\texttt{DllGetClassObject} qui retourne l'usine de classes du composant, un
serveur \emph{exe} doit manuellement enregistrer ses usines de classes en
appelant la fonction \texttt{CoRegisterClassObject} de l'API Win32. Rien de
bien compliqu\'e jusqu'\`a pr\'esent. Le probl\`eme majeur, c'est que le
syst\`eme n\'ecessite une librairie dynamique qui joue le r\^ole d'un 
\emph{proxy} (voir Fig. \ref{fig:outproc}). Le proxy communique avec une
instance de lui-m\^eme par le biais d'un protocole LPC (\emph{Local Procedure
Call}). Construire cette librairie dynamique est long et offre d'innombrables
possibilit\'es d'erreur.  

Le compilateur MIDL de Microsoft, qui prend comme source le fichier IDL
correspondant au composant fourni par le serveur, g\'en\`ere du code C
correspondant au proxy. Il est possible d'utiliser ce compilateur ainsi qu'un
compilateur C pour cr\'eer le proxy. Il est regrettable par ailleurs que cela
nous force \`a abandonner notre mod\`ele de garder la cr\'eation de programmes
Win32 compl\`etement centr\'ee sur des programmes SML. Nous \'etudions
pr\'esentement des m\'ethodes alternatives pour achever le m\^eme effet.

Nous notons en terminant qu'un serveur \emph{out-proc} peut \^etre utilis\'e
naturellement comme un serveur sur r\'eseau en utilisant le m\'ecanisme
DCOM. Ceci ne n\'ecessite aucune modification au code SML du serveur.

\section{Conclusion}

Nous d\'ecrivons dans cet article une infrastructure pour la
programmation Win32 sous le langage fonctionnel Standard ML. Un
prototype existe pour le compilateur Standard ML of New Jersey, et
sert comme base exp\'erimentale pour explorer diff\'erentes id\'ees.  

O\`u tout cela nous conduit-il? La majorit\'e des infrastructures pour la
programmation Win32 sous des langages tels que C et C++ introduisent des
librairies de haut-niveau qui r\'eduisent la complexit\'e de la
programmation, telles que les librairies de classes C++ MFC et OWL. Pour Standard
ML, une direction int\'eressante consiste en l'implantation d'une
librairie dans le style de eXene \cite{Gansner93}, une interface
graphique pour le syst\`eme X, qui 
utilise Concurrent ML \cite{Reppy91} pour traiter le parall\'elisme
implicite des interfaces graphiques. Un programme \`a long terme serait de
g\'en\'eraliser ces diff\'erentes librairies et de fournir une infrastructure
de programmation graphique uniforme sur diff\'erents syst\`emes, tout en
pr\'eservant les caract\'eristiques fondamentales des diff\'erents syst\`emes
(le \emph{look-and-feel}). Bien s\^ur, ceci n'addresse que le probl\`eme
de la programmation graphique, et laisse ouvert le probl\`eme d'utiliser des
composants de type COM (ou CORBA) d'une mani\`ere g\'en\'eralis\'ee.

Une remarque finale: notre infrastructure est bas\'e sur le compilateur
Standard ML of New Jersey, mais n'est certainement pas restreinte \`a ce
syst\`eme. En ajoutant \`a l'outil ML-IDL un \emph{backend} qui g\'en\`ere du
code pour la FFI de MLWorks ou de Ocaml \cite{UNSTABLE:Leroy96}, nous
pouvons fournir la m\^eme 
interface et les m\^emes capabilit\'es \`a ces diff\'erents syst\`emes. Par
ailleurs, d\^u \`a la nature des FFI de ces syst\`emes, nous devrons
probablement aussi g\'en\'erer du code C pour pouvoir implanter
l'\'equivalent de notre fonction \texttt{Wffi.funToAddr}.

\appendix

\section{Extrait de la description IDL de l'API Win32}
\label{app:idl}

\begin{alltt}\scriptsize
//
// IDL description of win32 API
//

sml_name ("W32");

typedef int INT;
typedef int HANDLE;
typedef int HRESULT;

typedef HANDLE HWND;
typedef HANDLE HDC;
typedef boolean BOOL;
typedef [string] char *STRING;
typedef [string] wchar_t *WSTRING;

typedef int *WNDPROC ([in] HWND hwnd, [in] INT msg, 
                      [in] INT wParam, [in] INT lParam);
        
typedef struct _WNDCLASSEX \{    // wc 
    UINT    cbSize; 
    int    style; 
    WNDPROC lpfnWndProc; 
    int     cbClsExtra; 
    int     cbWndExtra; 
    HANDLE  hInstance; 
    HANDLE   hIcon; 
    HANDLE hCursor; 
    HANDLE  hbrBackground; 
    STRING lpszMenuName; 
    STRING lpszClassName; 
    HANDLE   hIconSm; 
\} WNDCLASSEX; 

typedef struct tagPOINT \{ // pt 
    INT x; 
    INT y; 
\} POINT; 

const char *IDI_APPLICATION = "#32512";
const char *IDI_HAND = "#32513";
const char *IDI_QUESTION = "#32514";
const char *IDI_EXCLAMATION = "#32515";

const char *IDC_ARROW = "#32512";
const char *IDC_IBEAM = "#32513";
const char *IDC_WAIT = "#32514";
const char *IDC_CROSS = "#32515"; 

typedef enum \{  
 CS_VREDRAW = 1,
 CS_HREDRAW = 2,

 CW_USEDEFAULT = 0wx80000000,

 WS_OVERLAPPED =  0wx00000000,
 WS_POPUP =       0wx80000000,
 WS_CHILD =       0wx40000000,
 WS_MINIMIZE =    0wx20000000,
 WS_VISIBLE =     0wx10000000,

 ...

\} OPTS;

typedef enum \{
  SW_HIDE = 0,
  SW_SHOWNORMAL = 1,
  SW_NORMAL = 1,
  SW_SHOWMINIMIZED = 2,
  SW_SHOWMAXIMIZED = 3,
  SW_MAXIMIZE = 3,

  ...

\} CONSTS;

typedef enum \{
  WM_NULL          = 0wx0,
  WM_CREATE        = 0wx1,
  WM_DESTROY       = 0wx2,
  WM_MOVE          = 0wx3,
  WM_SIZE          = 0wx5,
  WM_SETFOCUS      = 0wx7,
  WM_PAINT         = 0wxf,
  WM_TIMER         = 0wx113,
  WM_HSCROLL       = 0wx114,
  WM_VSCROLL       = 0wx115,

  ...

\} WM;


[sml_source ("user32.dll")]
interface User \{

  // classes

  int RegisterClassExA ([in,ref] WNDCLASSEX *wndclass);

  BOOL UnregisterClassA ([in] STRING className, 
                         [in] HANDLE hInstance);

  // window handling

  HWND CreateWindowExA ([in] INT dwExStyle,
                        [in] STRING classname,
                        [in] STRING windowname,
                        [in] INT style,
                        [in] INT x,
                        [in] INT y,
                        [in] INT width,
                        [in] INT height,
                        [in] HWND parent,
                        [in] HANDLE menu,
                        [in] HANDLE hinstance,
                        [in] LPVOID param);

  BOOL ShowWindow ([in] HWND hwnd, 
                   [in] INT cmdshow);

  BOOL UpdateWindow ([in] HWND hwnd);

  // painting

  HDC BeginPaint ([in] HWND hwnd, 
                  [out,ref] PAINTSTRUCT *ps);

  BOOL EndPaint ([in] HWND hwnd, 
                 [in,ref] PAINTSTRUCT *ps);

  // icons

  HANDLE LoadIconA ([in] HANDLE h, 
                    [in] STRING name); 
  
\}



[sml_source ("gdi32.dll")]
interface Gdi \{

   BOOL LineTo ([in] HDC hdc, 
                [in] int nXEnd, 
                [in] int nYEnd);

   BOOL PolyLineTo ([in] HDC hdc,
                    [in,size_is (cPoints)] POINT *lppt,
                    [in] INT cPoints);

\}

\end{alltt}

\section{Extrait de la signature pour l'API Win32}
\label{app:sig}

\begin{alltt}\scriptsize

(**********************************************************************
 *
 *  This file was automatically generated by ml-idl
 *  (Tue Dec  8 09:41:24 1998)
 *
 **********************************************************************)

signature W32_SIG =
  sig
    (*
     * Pervasives
     *)
    type 'a pointer
    val null : 'a pointer
    val free : 'a pointer -> unit


    val IDI_APPLICATION : String.string
    val IDI_HAND : String.string
    val IDI_QUESTION : String.string
    val IDI_EXCLAMATION : String.string

    val IDC_ARROW : String.string
    val IDC_IBEAM : String.string
    val IDC_WAIT : String.string
    val IDC_CROSS : String.string

    type INT = Int32.int
    type HANDLE = Int32.int
    type HRESULT = Int32.int

    type HWND = HANDLE
    type HDC = HANDLE
    type BOOL = Bool.bool
    type STRING = String.string
    type WSTRING = String.string

    type WNDPROC = ((HWND * INT * INT * INT) -> Int32.int)

    datatype WNDCLASSEX = WNDCLASSEX of \{cbSize:UINT,
                                         style:Int32.int,
                                         lpfnWndProc:WNDPROC,
                                         cbClsExtra:Int32.int,
                                         cbWndExtra:Int32.int,
                                         hInstance:HANDLE,
                                         hIcon:HANDLE,
                                         hCursor:HANDLE,
                                         hbrBackground:HANDLE,
                                         lpszMenuName:STRING,
                                         lpszClassName:STRING,
                                         hIconSm:HANDLE\}

    datatype POINT = POINT of \{x:INT,y:INT\}

    datatype OPTS = CS_VREDRAW 
                  | CS_HREDRAW 
                  | CW_USEDEFAULT 
                  | WS_OVERLAPPED 
                  | WS_POPUP 
                  | WS_CHILD 
                  | WS_MINIMIZE 
                  | WS_VISIBLE 
                  | ...
    structure OPTS : sig
      val toInt : OPTS -> Int32.int
      val fromInt : Int32.int -> OPTS option
    end

    datatype CONSTS = SW_HIDE 
                    | SW_SHOWNORMAL 
                    | SW_NORMAL 
                    | SW_SHOWMINIMIZED 
                    | SW_SHOWMAXIMIZED 
                    | SW_MAXIMIZE 
                    | ...
    structure CONSTS : sig
      val toInt : CONSTS -> Int32.int
      val fromInt : Int32.int -> CONSTS option
    end

    datatype WM = WM_NULL 
                | WM_CREATE 
                | WM_DESTROY 
                | WM_MOVE 
                | WM_SIZE 
                | WM_SETFOCUS 
                | WM_PAINT 
                | WM_TIMER 
                | WM_HSCROLL 
                | WM_VSCROLL 
                | ...
    structure WM : sig
      val toInt : WM -> Int32.int
      val fromInt : Int32.int -> WM option
    end

    structure User : sig
      val RegisterClassExA : WNDCLASSEX -> Int32.int
      val UnregisterClassA : (STRING * HANDLE) -> BOOL
      val CreateWindowExA : (INT * STRING * STRING * INT * 
                             INT * INT * INT * INT * HWND * 
                             HANDLE * HANDLE * LPVOID) -> HWND
      val ShowWindow : (HWND * INT) -> BOOL
      val UpdateWindow : HWND -> BOOL
      val BeginPaint : HWND -> (PAINTSTRUCT * HDC)
      val EndPaint : (HWND * PAINTSTRUCT) -> BOOL
      val LoadIconA : (HANDLE * STRING) -> HANDLE
    end

    structure Gdi : sig
      val LineTo : (HDC * Int32.int * Int32.int) -> BOOL
      val PolyLineTo : (HDC * POINT list * INT) -> BOOL
    end

  end

\end{alltt}

\section{Exemple de programme Win32}
\label{app:exemple}

\begin{alltt}\scriptsize

structure BounceWin = struct
  
  val zero = 0 : Int32.int
    
  val ballTimer = 2 : Int32.int
    
  val moveRate = 10 : Int32.int
  val timerRate = 20 : Int32.int
    
  val hInstance = ref (zero)
    
  val cxClient = ref (zero)
  val cyClient = ref (zero)
  val xCenter = ref (zero)
  val yCenter = ref (zero)
  val cxTotal = ref (zero)
  val cyTotal = ref (zero)
  val cxRadius = ref (zero)
  val cyRadius = ref (zero)
  val cxMove = ref (zero)
  val cyMove = ref (zero)
  val hBitmap = ref (zero)
    
  fun destroy (hwnd) = 
    (W32.User.KillTimer (hwnd,ballTimer);
     if (!hBitmap <> 0)
       then ignore (W32.Gdi.DeleteObject (!hBitmap))
     else ();
       W32.User.PostQuitMessage (0); 
     zero)
    
  fun size (hwnd, xsize, ysize) = 
    (cxClient := xsize;
     cyClient := ysize;
     xCenter := xsize div 2;
     yCenter := ysize div 2;
     cxMove := moveRate;
     cyMove := moveRate;
     cxTotal := 158;  (* from PAINT *)
     cyTotal := 131; 
     cxRadius := 118 div 2;
     cyRadius := 90 div 2;
     if (!hBitmap <> 0)
       then ignore (W32.Gdi.DeleteObject (!hBitmap))
     else ();
     hBitmap := W32.User.LoadImageA 
                    (0,"smlnj.bmp",
                     W32.CONSTS.toInt
                     (W32.IMAGE_BITMAP),
                     0,0,W32U.or [W32.LR_LOADFROMFILE]);
     zero)
    
  fun timerBall (hwnd) = 
    if (!hBitmap = 0) then zero
    else let 
      val hdc = W32.User.GetDC (hwnd)
      val hdcMem = W32.Gdi.CreateCompatibleDC (hdc)
    in
      W32.Gdi.SelectObject (hdcMem, !hBitmap);
      W32.Gdi.BitBlt (hdc, 
                      (!xCenter) - (!cxTotal) div 2,
                      (!yCenter) - (!cyTotal) div 2,
                      !cxTotal, !cyTotal, 
                      hdcMem, 0,0,
                      W32.CONSTS.toInt (W32.SRCCOPY));
      W32.User.ReleaseDC (hwnd, hdc);
      W32.Gdi.DeleteDC (hdcMem);
      xCenter := (!xCenter) + (!cxMove);
      yCenter := (!yCenter) + (!cyMove);
      if ((!xCenter) + (!cxRadius) >= (!cxClient)) orelse
        ((!xCenter) - (!cxRadius) <= 0)
        then cxMove := ~(!cxMove) 
      else ();
      if ((!yCenter) + (!cyRadius) >= (!cyClient)) orelse
        ((!yCenter) - (!cyRadius) <= 0)
        then cyMove := ~(!cyMove)
      else ();
        zero
    end
  
  fun timer (hwnd, timerID) = 
    if (timerID = ballTimer)
      then timerBall (hwnd)
    else zero
                                
                                
  fun create (hwnd) = let 
    val hdc = W32.User.GetDC (hwnd)
  in
    W32.User.ReleaseDC (hwnd,hdc);
    W32.User.SetTimer (hwnd, ballTimer, timerRate, NONE);
    zero
  end

  and wndproc (hwnd, Msg, wParam, lParam) = 
    case (W32.WM.fromInt (Msg))
      of SOME (W32.WM_CREATE) => create (hwnd)
       | SOME (W32.WM_SIZE) => 
          size (hwnd, W32U.loWord (lParam), W32U.hiWord (lParam))
       | SOME (W32.WM_DESTROY) => destroy (hwnd)
       | SOME (W32.WM_TIMER) => timer (hwnd, wParam)
       | _ => W32.User.DefWindowProcA (hwnd,Msg,wParam,lParam)
        
  fun winmain (hinstance,_) = let
    val _ = hInstance := hinstance
    val szAppName = "BouncingSMLNJ"
    val hicon = W32.User.LoadIconA (0, W32.IDI_APPLICATION)
    val hcursor = W32.User.LoadCursorA (0, W32.IDC_ARROW)
    val hbrush = W32.Gdi.GetStockObject (W32.CONSTS.toInt (W32.WHITE_BRUSH))
    val wndclassex = W32.WNDCLASSEX 
                       \{cbSize = 48,
                        style = W32U.or [W32.CS_HREDRAW,
                                         W32.CS_VREDRAW],
                        lpfnWndProc = wndproc,
                        cbClsExtra = 0,
                        cbWndExtra = 0,
                        hInstance = hinstance,
                        hIcon = hicon,
                        hCursor = hcursor,
                        hbrBackground = hbrush,
                        lpszMenuName = "",
                        lpszClassName = szAppName,
                        hIconSm = hicon\}
    val _ = W32.User.RegisterClassExA (wndclassex)  
    val hwnd = W32.User.CreateWindowExA 
                         (0,
                          szAppName,
                          "Bouncing SML/NJ",
                          W32U.or [W32.WS_OVERLAPPEDWINDOW],
                          W32U.or [W32.CW_USEDEFAULT],
                          W32U.or [W32.CW_USEDEFAULT],
                          500,
                          300,
                          0,
                          0,
                          hinstance,
                          W32.null)
    fun f () = (W32.User.ShowWindow (hwnd, 1);
                W32.User.UpdateWindow (hwnd);
                (* the following call is required because creating
                 * the window from the interactive loop seems to 
                 * not put it on top
                 *)
                W32.User.SetForegroundWindow (hwnd);
                W32U.simpleMsgLoop ())
  in
    W32U.checkHwnd (hwnd,f);
    W32.User.UnregisterClassA (szAppName, hinstance);
    ()
  end

  fun doit () = RunW32.doit (winmain)
    
end

\end{alltt}

\end{document}